\documentclass[twocolumn,tightenlines,showpacs, floatfix,prd,superscriptaddress,showpacs,amssymb,amsmath,amsfonts,aps,altaffilletter]{revtex4}

\usepackage{graphicx}

\usepackage{bm}

\begin{document}

\title{First LIGO search for gravitational wave bursts from cosmic
  (super)strings}

%
%
%

\newcommand*{\AG}{Albert-Einstein-Institut, Max-Planck-Institut f\"{u}r Gravitationsphysik, D-14476 Golm, Germany}
\affiliation{\AG}
\newcommand*{\AH}{Albert-Einstein-Institut, Max-Planck-Institut f\"{u}r Gravitationsphysik, D-30167 Hannover, Germany}
\affiliation{\AH}
\newcommand*{\AU}{Andrews University, Berrien Springs, MI 49104 USA}
\affiliation{\AU}
\newcommand*{\AN}{Australian National University, Canberra, 0200, Australia}
\affiliation{\AN}
\newcommand*{\CH}{California Institute of Technology, Pasadena, CA  91125, USA}
\affiliation{\CH}
\newcommand*{\CA}{Caltech-CaRT, Pasadena, CA  91125, USA}
\affiliation{\CA}
\newcommand*{\CU}{Cardiff University, Cardiff, CF24 3AA, United Kingdom}
\affiliation{\CU}
\newcommand*{\CL}{Carleton College, Northfield, MN  55057, USA}
\affiliation{\CL}
\newcommand*{\CS}{Charles Sturt University, Wagga Wagga, NSW 2678, Australia}
\affiliation{\CS}
\newcommand*{\CO}{Columbia University, New York, NY  10027, USA}
\affiliation{\CO}
\newcommand*{\ER}{Embry-Riddle Aeronautical University, Prescott, AZ   86301 USA}
\affiliation{\ER}
\newcommand*{\EU}{E\"{o}tv\"{o}s University, ELTE 1053 Budapest, Hungary}
\affiliation{\EU}
\newcommand*{\HC}{Hobart and William Smith Colleges, Geneva, NY  14456, USA}
\affiliation{\HC}
\newcommand*{\IA}{Institute of Applied Physics, Nizhny Novgorod, 603950, Russia}
\affiliation{\IA}
\newcommand*{\IU}{Inter-University Centre for Astronomy  and Astrophysics, Pune - 411007, India}
\affiliation{\IU}
\newcommand*{\HU}{Leibniz Universit\"{a}t Hannover, D-30167 Hannover, Germany}
\affiliation{\HU}
\newcommand*{\CT}{LIGO - California Institute of Technology, Pasadena, CA  91125, USA}
\affiliation{\CT}
\newcommand*{\LO}{LIGO - Hanford Observatory, Richland, WA  99352, USA}
\affiliation{\LO}
\newcommand*{\LV}{LIGO - Livingston Observatory, Livingston, LA  70754, USA}
\affiliation{\LV}
\newcommand*{\LM}{LIGO - Massachusetts Institute of Technology, Cambridge, MA 02139, USA}
\affiliation{\LM}
\newcommand*{\LU}{Louisiana State University, Baton Rouge, LA  70803, USA}
\affiliation{\LU}
\newcommand*{\LE}{Louisiana Tech University, Ruston, LA  71272, USA}
\affiliation{\LE}
\newcommand*{\LL}{Loyola University, New Orleans, LA 70118, USA}
\affiliation{\LL}
\newcommand*{\MT}{Montana State University, Bozeman, MT 59717, USA}
\affiliation{\MT}
\newcommand*{\MS}{Moscow State University, Moscow, 119992, Russia}
\affiliation{\MS}
\newcommand*{\ND}{NASA/Goddard Space Flight Center, Greenbelt, MD  20771, USA}
\affiliation{\ND}
\newcommand*{\NA}{National Astronomical Observatory of Japan, Tokyo  181-8588, Japan}
\affiliation{\NA}
\newcommand*{\NO}{Northwestern University, Evanston, IL  60208, USA}
\affiliation{\NO}
\newcommand*{\RI}{Rochester Institute of Technology, Rochester, NY  14623, USA}
\affiliation{\RI}
\newcommand*{\RA}{Rutherford Appleton Laboratory, HSIC, Chilton, Didcot, Oxon OX11 0QX United Kingdom}
\affiliation{\RA}
\newcommand*{\SJ}{San Jose State University, San Jose, CA 95192, USA}
\affiliation{\SJ}
\newcommand*{\SM}{Sonoma State University, Rohnert Park, CA 94928, USA}
\affiliation{\SM}
\newcommand*{\SE}{Southeastern Louisiana University, Hammond, LA  70402, USA}
\affiliation{\SE}
\newcommand*{\SO}{Southern University and A\&M College, Baton Rouge, LA  70813, USA}
\affiliation{\SO}
\newcommand*{\SA}{Stanford University, Stanford, CA  94305, USA}
\affiliation{\SA}
\newcommand*{\SR}{Syracuse University, Syracuse, NY  13244, USA}
\affiliation{\SR}
\newcommand*{\PU}{The Pennsylvania State University, University Park, PA  16802, USA}
\affiliation{\PU}
\newcommand*{\UM}{The University of Melbourne, Parkville VIC 3010, Australia}
\affiliation{\UM}
\newcommand*{\MI}{The University of Mississippi, University, MS 38677, USA}
\affiliation{\MI}
\newcommand*{\SF}{The University of Sheffield, Sheffield S10 2TN, United Kingdom}
\affiliation{\SF}
\newcommand*{\TA}{The University of Texas at Austin, Austin, TX 78712, USA}
\affiliation{\TA}
\newcommand*{\TC}{The University of Texas at Brownsville and Texas Southmost College, Brownsville, TX  78520, USA}
\affiliation{\TC}
\newcommand*{\TR}{Trinity University, San Antonio, TX  78212, USA}
\affiliation{\TR}
\newcommand*{\BB}{Universitat de les Illes Balears, E-07122 Palma de Mallorca, Spain}
\affiliation{\BB}
\newcommand*{\UA}{University of Adelaide, Adelaide, SA 5005, Australia}
\affiliation{\UA}
\newcommand*{\BR}{University of Birmingham, Birmingham, B15 2TT, United Kingdom}
\affiliation{\BR}
\newcommand*{\FA}{University of Florida, Gainesville, FL  32611, USA}
\affiliation{\FA}
\newcommand*{\GU}{University of Glasgow, Glasgow, G12 8QQ, United Kingdom}
\affiliation{\GU}
\newcommand*{\MD}{University of Maryland, College Park, MD 20742 USA}
\affiliation{\MD}
\newcommand*{\AM}{University of Massachusetts - Amherst, Amherst, MA 01003, USA}
\affiliation{\AM}
\newcommand*{\MU}{University of Michigan, Ann Arbor, MI  48109, USA}
\affiliation{\MU}
\newcommand*{\MN}{University of Minnesota, Minneapolis, MN 55455, USA}
\affiliation{\MN}
\newcommand*{\OU}{University of Oregon, Eugene, OR  97403, USA}
\affiliation{\OU}
\newcommand*{\RO}{University of Rochester, Rochester, NY  14627, USA}
\affiliation{\RO}
\newcommand*{\SL}{University of Salerno, 84084 Fisciano (Salerno), Italy}
\affiliation{\SL}
\newcommand*{\SN}{University of Sannio at Benevento, I-82100 Benevento, Italy}
\affiliation{\SN}
\newcommand*{\SH}{University of Southampton, Southampton, SO17 1BJ, United Kingdom}
\affiliation{\SH}
\newcommand*{\SC}{University of Strathclyde, Glasgow, G1 1XQ, United Kingdom}
\affiliation{\SC}
\newcommand*{\WA}{University of Western Australia, Crawley, WA 6009, Australia}
\affiliation{\WA}
\newcommand*{\UW}{University of Wisconsin-Milwaukee, Milwaukee, WI  53201, USA}
\affiliation{\UW}
\newcommand*{\WU}{Washington State University, Pullman, WA 99164, USA}
\affiliation{\WU}

\author{}    \affiliation{\GU}    
\author{B.~P.~Abbott}    \affiliation{\CT}    
\author{R.~Abbott}    \affiliation{\CT}    
\author{R.~Adhikari}    \affiliation{\CT}    
\author{P.~Ajith}    \affiliation{\AH}    
\author{B.~Allen}    \affiliation{\AH}  \affiliation{\UW}  
\author{G.~Allen}    \affiliation{\SA}    
\author{R.~S.~Amin}    \affiliation{\LU}    
\author{S.~B.~Anderson}    \affiliation{\CT}    
\author{W.~G.~Anderson}    \affiliation{\UW}    
% ...
 \collaboration{The LIGO Scientific Collaboration, http://www.ligo.org}
 \noaffiliation
\noaffiliation
\author{F.~Robinet}\affiliation{LAL, Univ Paris-Sud, CNRS/IN2P3,
Orsay, France}

%
%

\date{\today}

\begin{abstract}
  We report on a matched-filter search for gravitational wave bursts
  from cosmic string cusps using LIGO data from the fourth science run
  (S4) which took place in February and March 2005.  No gravitational
  waves were detected in 14.9 days of data from times when all three
  LIGO detectors were operating.  We interpret the result in terms of
  a frequentist upper limit on the rate of gravitational wave bursts
  and use the limits on the rate to constrain the parameter space
  (string tension, reconnection probability, and loop sizes) of cosmic
  string models.
\end{abstract}


\maketitle

\section{Introduction}

Cosmic strings are one-dimensional topological defects that can form
during phase transitions in the early universe
\cite{kibble76,alexbook}. Topological defect formation is generic in
grand unified theories (GUTs), and cosmic string production
specifically is generic in supersymmetric
GUTs~\cite{Jeannerot:2003qv}. In string theory motivated cosmological
models, cosmic strings may also form (and are referred to as cosmic
superstrings to differentiate them from strings formed 
  in phase transitions)~\cite{Jones:2002cv,%
  Sarangi:2002yt,Dvali:2003zj,Jones:2003da,Copeland:2003bj,%
  Jackson:2004zg,Tye:2005fn,Copeland:2006if,Leblond:2007tf,%
  Rajantie:2007hp}.
Cosmic strings and superstrings may produce a variety of
astrophysical signatures including gamma ray bursts~\cite{bands},
ultra-high energy cosmic rays~\cite{berez},
magnetogenesis~\cite{Battefeld:2007qn}, microlensing, strong and weak
lensing~\cite{Chernoff:2007pd,Kuijken:2007ma,Gasparini:2007jj,Christiansen:2008vi,Dyda:2007su},
radio bursts~\cite{Vachaspati:2008su}, effects on the cosmic 21~cm
power spectrum~\cite{Khatri:2008zw}, effects on the cosmic microwave
background (CMB) at small angular
scales~\cite{Fraisse:2007nu,Pogosian:2008am}, and effects on the CMB
polarization \cite{Baumann:2008aj}.

Cosmic strings and superstrings can also produce powerful bursts of
gravitational waves~\cite{firstcusps, DV0,DV1,DV2,SCMMCR}.  The most
potent bursts are produced at regions of string called cusps which acquire large
Lorentz boosts.  The formation of cusps on cosmic strings
is generic and cusp gravitational waveforms are simple and
robust~\cite{kenandi,damour}. The large mass per unit length of cosmic
strings combined with the large Lorentz boost may result in signals
detectable by Earth-based  interferometric gravitational wave detectors
such as LIGO \cite{ligoref} and Virgo \cite{virgoref}. Thus,
gravitational waves may provide a powerful probe of early universe
physics.

The LIGO detector network is comprised of three laser interferometers.
Two of them are located at the Hanford, WA site: a four-kilometer arm
instrument referred to as H1, and a two-kilometer arm instrument
referred to as H2. A second four-kilometer interferometer located at
the Livingston, LA site, is referred to as L1.  LIGO's fourth
science run (S4) took place between February 22, 2005 and March 23,
2005.  The configuration of the LIGO instruments during the fourth
science run (S4), is described in~\cite{sigg}. The sensitivity of this
run was significantly better than that of previous runs: at the low
frequencies relevant to this search, close to a factor of ten more
sensitive than the previous science run S3, though still about a
factor of two less sensitive than LIGO's most recent science run
(S5), which was at design sensitivity.

In this work we report on the results of a matched-filter search for
bursts from cosmic string cusps performed on 14.9 days of S4 data. We
implement the data analysis methods described in~\cite{SCMMCR} using a
simple triple coincidence scheme.  No gravitational waves were
detected and we interpret the result in terms of a frequentist upper
limit on the rate using the loudest event
technique~\cite{loudestevent}. We use the upper limit on the rate to
constrain the parameter space of cosmic strings models. The
sensitivity of the LIGO instruments during the S4 run does not allow us to
place constraints as tight as the indirect bounds from Big Bang
Nucleosynthesis~\cite{SMC}. In the future, however, we expect our sensitivity to
surpass these limits for large areas of cosmic string model
parameter space.

In Sec.~\ref{DA} we discuss data selection, data analysis techniques,
and describe the analysis pipeline.  In Sec.~\ref{FGBG} we describe
the computation of the rate of accidental events we expect to survive
the thresholds and consistency checks of the pipeline (the so-called
background), and compare it to the events that made it to the end of
the pipeline in our search (the so-called foreground). In
Sec.~\ref{EFF} we show how we estimate the sensitivity of the analysis
using simulated gravitational-wave signals.  We compute the
efficiency of our pipeline, the fraction of simulated signals that we
detect, as a function of the strength of the signals. In
Sec.~\ref{cosmo} we show how to estimate the rate of burst events we
expect, the effective rate, using the efficiency curves and the
cosmological rate of events.  We show the constraints our data place
on the parameter space of cosmic string models. We conclude in
Sec.~\ref{concl}.

\section{Data Analysis}
\label{DA}

\subsection{Data selection and conditioning}

All available S4 science data when all three instruments were
operating (triple-coincident data) were used except for
periods
\begin{enumerate}
\item with overflows in the error signal digitizer,
\item when airplanes flew over the detector sites,
\item thirty seconds prior to loss of lock (loss of resonance of the
  Fabry-Perot cavities in the arms) of any instrument,
\item of excessive wind,
\item of excessive seismic activity, and
\item with calibration uncertainties larger than 10\%.
\end{enumerate} 
The total time of triple coincident data available after
these cuts is 14.9 days. Calibration of the data used in this analysis
was performed in the time domain~\cite{htpaper}. The data were
high-pass filtered near 30~Hz to remove un-necessary low frequency
content, and down-sampled from the original LIGO sampling rate of
16384~Hz to 4096~Hz.

\subsection{Matched-filters and templates}

For each of the three LIGO instruments we then performed a
matched-filter search on this data for gravitational bursts from
cosmic string cusps, i.e.  linearly polarized signals of the 
form~\cite{DV0}
\begin{equation}
h_+(f) = B f^{-4/3} \Theta(f_h-f)\,\Theta(f-f_l).
\label{eq:hf1}
\end{equation}
The amplitude of the cusp
waveform is $B \sim G\mu L^{2/3}/(c^3r)$, where $G$ is Newton's constant,
$\mu$ is the mass per unit length of the string, $L$ is the size of
the feature on the string that produces the cusp, and $r$ is the
distance between the cusp and the point of observation.  In natural
units $G\mu/c^2$ can be thought of as the dimensionless mass per unit
length, or tension, of cosmic strings. The size $L$ of the feature on
the string that produces the cusp also determines the low frequency
cutoff $f_l$.  Since $L$ is expected to be cosmological, for example
the size of a cosmic string loop, the low frequency cutoff of {\it
  detectable} radiation is determined by the low frequency behavior of
the instruments: for the LIGO instruments by seismic noise. The high
frequency cutoff depends on the angle $\theta$ between the line of
sight and the direction of the cusp. It is given by $f_h \sim
2c/(\theta^3 L)$ and can be arbitrarily large (up to the inverse of
the light crossing time of the width of strings).

Following~\cite{SCMMCR} for our templates we take
\begin{equation}
\tau(f) = f^{-4/3} \Theta(f_h-f)\,\Theta(f-f_l),
\label{eq:templ1}
\end{equation}
so that $h(f) = A\, \tau(f)$. We can normalize our templates by defining
the detector-noise-weighted inner product~\cite{cutlerflanagan} in
terms of the two frequency series $x(f)$ and $y(f)$ as
\begin{eqnarray}
(x|y) \equiv 4 \Re \int_0^\infty df \,\frac{x(f)y^*(f)}{S_h(f)}.
\label{eq:dorprod1}
\end{eqnarray}
Here $S_h(f)$ is the single-sided spectral density defined by $\langle
n(f) n^*(f')\rangle = \frac{1}{2} \delta(f-f') S_h(f)$ and $n(f)$ is
the Fourier transform of the detector noise.  We take the inner
product of a template with itself to be $\sigma^2=(\tau|\tau)$ and define the
normalized template ${\hat \tau} \equiv \tau/\sigma$, so that $({\hat
  \tau}|{\hat \tau})=1$.  

The calibrated output of an interferometer can be written as
\begin{equation}
s(t)=n(t)+h(t), 
\end{equation}
where $n(t)$ is the instrumental noise, and $h(t)=F_+h_+(t)$ the
gravitational wave signal.  The antenna pattern response function to
$+$-polarized gravitational waves, $F_+$, is a function of the
sky-location of the cusp and the polarization angle. The signal to
noise ratio (SNR) is defined in terms of the inner product as $\rho
\equiv (s|{\hat \tau})$.  For the case of Gaussian noise and in the
absence of a signal the SNR is a Gaussian variable with zero mean and
unit variance. In the presence of a signal of amplitude $A$, the
signal to noise ratio is a Gaussian random variable, with mean
$A\sigma$ and unit variance. Since the average SNR is $\langle \rho
\rangle = A \sigma$, for a particular realization of the measured SNR
$\rho$ we can identify a signal amplitude
\begin{equation}
 A = \rho/\sigma,
\end{equation}
which has an average value $\langle A \rangle = F_+ B$.

In our search we set $f_l=50$~Hz to be our low frequency cutoff. Due to
the low frequency behavior of our instruments, 
a negligible SNR would be gained by including
frequencies lower than $50$~Hz. We look for
signals with high frequency cutoffs $f_h$ in the range $75$~Hz-$2048$~Hz.  The
sensitivity of the instruments is such that very little is lost by
limiting the search to signals with high frequency cutoffs above 75~Hz
and below 2048~Hz. The only template parameter is the high frequency
cutoff $f_h$ and the template bank (the set of templates the determine the
signals we search for) is constructed iteratively by computing the
overlap between adjacent templates~\cite{SCMMCR}.  The maximum
fractional loss of signal to noise is set to $0.05$ and along with the
spectrum determines the spacing between the high frequency cutoffs of
the different templates.  The spectrum $S_h(f)$ is estimated using the
median-mean method~\cite{Allen:2005fk} which is fairly robust against
non-stationarities in the data, including loud simulated signal
injections.

\subsection{Trigger generation}

To produce our trigger data we proceed as follows. We apply the
matched-filter for each template and all possible arrival times using
fast Fourier transform convolution (as described in \cite{SCMMCR}).
This procedure results in a time series for the SNR sampled at 4096~Hz
for every template. In each of these time series we search for
clusters of values above the threshold $\rho_{\text{th}}=4$ which we
identify as triggers. For each trigger we determine
\begin{itemize} 
\item[(a)] the SNR $\rho$ of the trigger (the maximum SNR of the cluster), 
\item[(b)] the peak time of the trigger (the location in time of the
  trigger SNR),
\item[(c)] the start time of the trigger (the first value above threshold
  in the cluster),  
\item[(d)] the duration of the trigger (the length of the cluster), 
\item[(e)] the high frequency cutoff of the template, and, 
\item[(f)] the amplitude of the trigger $A$, given by $A=\rho/\sigma$, where
  $\sigma$ is the template normalization.
\end{itemize}
When several templates result in triggers that occur within a time of
$0.1$~s we select the trigger with the largest SNR within that time
window. 

We apply this procedure to the data sets of the three LIGO
interferometers to produce a list of triggers for each instrument.  

\subsection{Trigger consistency checks}

To reduce the rate of events unassociated with gravitational waves
(noise induced events) we demand that the peak times of triggers in
each instrument be coincident in time with triggers in the other two
instruments.  The time window used for H1-H2 events is $2$ms. This
coincidence window allows for calibration uncertainties as well as
shifts in the peak times of triggers induced by fluctuations in the
noise. For coincidence between events in either of the two Hanford
instruments with events in the Livingston interferometer a $12$ms
coincidence window is used.  This allows for the maximum light travel
time between sites of $10$ms along with calibration uncertainties and
shifting of the peak location due to noise. We require strict triple
coincidence: in order for an event in one interferometer to survive it
must be coincident with events in the other two instruments and those
two events must also be coincident.

Additionally, we impose a symmetric consistency check on the
amplitudes of H1 and H2 coincident events \cite{SCMMCR}. In
particular, for H1 events we demand that
\begin{equation}
\frac{\left|A_{\text{H1}}-A_{\text{H2}}\right|} {A_{\text{H1}}}
<  \left(\frac{\delta}{\rho_{\text{H1}}} + \kappa \right),
\label{eq:Damp2}
\end{equation}
along with an analogous requirement for H2 events.  Here, $\delta$ is
the number of standard deviations of amplitude difference we allow and
$\kappa$ is an additional fractional difference that accounts for
other sources of uncertainty such as the calibration.  We
conservatively set $\delta = 3$ and $\kappa=0.5$. The purpose of this
loose cut on the amplitude of events is to eliminate large SNR events
seen in one instrument but not in the other. These events are not due
to gravitational waves but rather to instrumental glitches in the data
streams. It is worth pointing out that because each instrument has its
own antenna response factor $F_+$ that depends on the orientation and
direction of the gravitational wave, the H1-H2 amplitude consistency
test cannot be applied to H1-L1 or H2-L1. Instruments that are not
co-located may have different values of $F_+$ and therefore the
measured amplitudes $A$ could differ significantly.

Both time coincidence and amplitude consistency checks reduce the rate
of events in each instrument from about $1$~Hz to about $10$~$\mu$Hz.  To
simplify the analysis in the following we only use the results and
statistics for the H1 events.

\begin{figure}[t]
\resizebox*{\columnwidth}{!}{\includegraphics{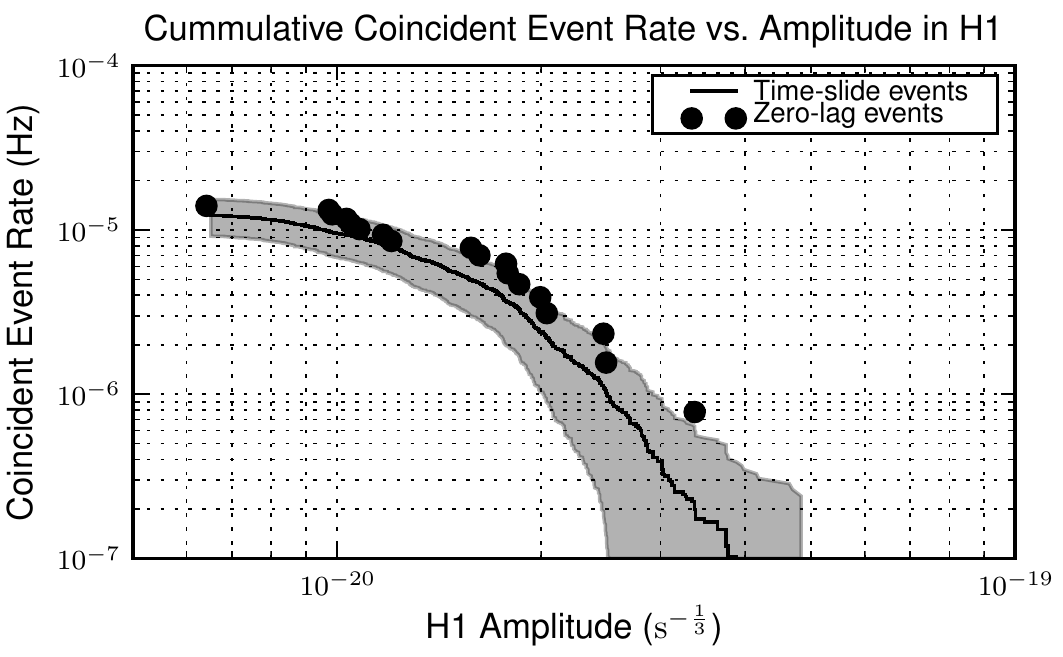}}
\caption{Plot of the cumulative rate of events as a function of the
  amplitude for both foreground events (filled circles), as well as average
  number of events found in the time shifted data (stair steps).  The shaded
  region corresponds to $1$-$\sigma$ variations measured in the time
  shifts.}
\label{fig:fgH1}
\end{figure}

\section{Foreground and Background}
\label{FGBG}

To estimate the rate of accidental coincidences that survive our
consistency checks, the so-called background, we time shift the
trigger sets relative to one another and look for coincident events.
We do not perform time shifts on the two Hanford trigger sets.
Environmental disturbances are known to cause correlations between
triggers in both interferometers and the effect of time shifting the
two Hanford trigger sets would be to underestimate the background.  We
therefore take the double coincident H1 and H2 triggers (which already satisfy
the amplitude consistency check) and time-shift them relative to the
L1 trigger set. This amounts to treating the two Hanford instruments as a
single trigger generator, on the same footing as L1 but with a much
smaller trigger rate. For each trigger in each time shift we then
demand the the first consistency criterion be satisfied, namely, that
each Hanford trigger peak be within $12$ms of a Livingston trigger
peak.  We performed 100 time shifts, with Livingston triggers shifted
by approximately $1.79$~s, the total time shift ranging from $-89.3$~s
to $89.3$~s.  The time shifts are sufficiently large that coincident
events cannot result from gravitational wave bursts from cosmic string
cusps.

Figure~\ref{fig:fgH1} summarizes the results of this procedure for the
H1 trigger set.  We plot the cumulative rate of events for both
foreground (unshifted) events (filled circles), as well as the average
rate of events found in the time shifted data (stair-steps) binned in
amplitude.  The shaded region corresponds to $1$-$\sigma$
uncertainties computed from the variations in the number of events
found in the time shifted data.

The loudest H1 event has an amplitude of $A^L = 3.4 \times
10^{-20}$~s$^{-1/3}$. There are no foreground events which deviate
significantly from the time-slides, and a Kolmogorov-Smirnov test
confirms that the foreground and background distributions are
consistent at the 77\% confidence level. We
therefore conclude that no gravitational waves have been detected in
this search.

\section{Efficiency}
\label{EFF}

To determine our sensitivity and construct an upper limit we injected
over 7400 simulated cusp signals into our data set and performed a
search identical to the one described above.

The distribution of high frequency cutoffs $f_h$ for the injected
signals is $dN \propto f_h^{-5/3}df_h$, appropriate for the cusp
signals we are seeking \cite{SCMMCR}.  The lowest high frequency
cutoff injected is $f_* = 75$~Hz, coinciding with the lowest high
frequency cutoff of our templates. The amplitudes are distributed
logarithmically between $B=6\times 10^{-21}$~s$^{-1/3}$ and
$B=10^{-17}$~s$^{-1/3}$ spanning the range of detectability. The sources
are placed isotropically in the sky with sufficient separation in time
so as not to unduly bias the spectrum estimate needed to perform the
matched-filter.

An injection is found if its peak time lies between the start time and
the end time of an H1 triple coincident trigger. We record the
recovered amplitude of the injection, which is different (typically
smaller) than the injected amplitude because of antenna pattern
effects as well as noise induced fluctuations.

\begin{figure}[t]
\resizebox*{\columnwidth}{!}{\includegraphics{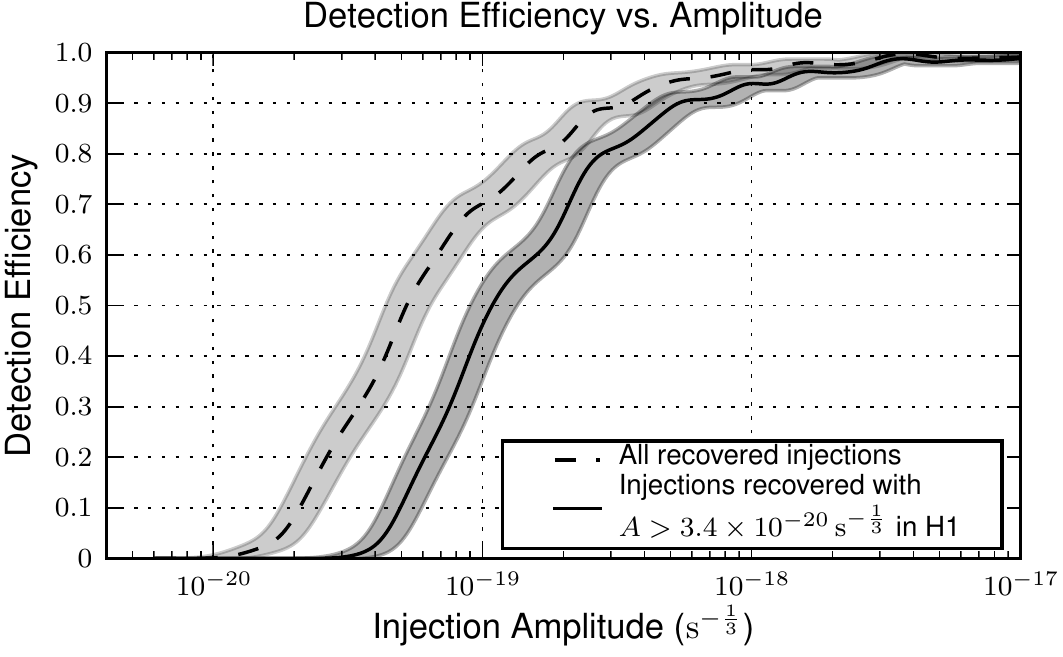}}
\caption{Plot of the detection efficiency $\epsilon(B)$ (fraction of events
  detected) as a function of the injected amplitude $B$. The solid curve
  corresponds to injections with recovered amplitudes $A$ larger than our
  loudest event $A^L = 3.4 \times 10^{-20}$~s$^{-1/3}$, and the dashed
  curve to all recovered injections. The shaded regions indicate our
  uncertainty in the efficiencies and amplitudes that result from
  counting, binning and calibration systematics.}
\label{fig:effH1}
\end{figure}

The result of our injection run is summarized in Fig.~\ref{fig:effH1}.
We plot the efficiency $\epsilon(B)$ (the fraction of injections
detected in the triple coincident H1 trigger set) as a function of the
injected amplitude $B$ for recovered H1 amplitudes $A$ greater than
our loudest event $A^L = 3.4 \times 10^{-20}$~s$^{-1/3}$(solid line),
and for \emph{any} recovered amplitudes (dashed line). The shaded
regions indicate our uncertainties in the efficiencies and amplitudes.
The procedure by which these curves and their uncertainties are
produced are described below in more detail.

A useful measure of our sensitivity is the injected amplitude at which
we recover half of our injections.  For all recovered injections this
amplitude is 
\begin{equation}
B_{50\%}=\left( 5.2 \pm 0.9 \right) \times 10^{-20}{\rm s}^{-1/3},
\label{eq:A50sens}
\end{equation}
and for those recovered with amplitudes above our loudest event above
it is 
\begin{equation}
B^L_{50\%}=\left( 1.1 \pm 0.2 \right)\times 10^{-19}{\rm s}^{-1/3}.
\label{eq:A50UL}
\end{equation}
These numbers are consistent with our expectations. In~\cite{SCMMCR} a
sensitivity estimate was made for Initial LIGO $B^{{\rm LIGO}}_{50\%}
\approx 10^{-20}$~s$^{-1/3}$.  The current search is somewhat less
sensitive for two reasons. First, data from the S4 run is about a
factor of two less sensitive than Initial LIGO. Second, we demand
coincidence with events in the H2 interferometer which is about a
factor of two less sensitive than H1. Together these account for a
factor of about four leaving only about a $30\%$ discrepancy between
the rough Initial LIGO sensitivity estimate made in~\cite{SCMMCR} 
and Eq.~(\ref{eq:A50sens}).

As stated above, the software injections are generated randomly, with
injected amplitudes that are uniformly distributed in their logarithm.
Individually, each injection is either found or missed.  To estimate
the probability of injections with a given amplitude being recovered
we used a sliding window to count the number of software injections
that were made and recovered within an interval around that amplitude.
The window used was Gaussian in the logarithm of the injected
amplitudes.  Choosing different widths for the window will yield
qualitatively equivalent but quantitatively different efficiency
curves.  As pointed out above, a useful measure of our sensitivity is
\(B_{50\%}\), and we chose the width of the Gaussian window to
minimize the uncertainty in \(B_{50\%}\).

Three uncertainties are associated with the efficiency curve.
First, because at each point the value of the efficiency has been
measured by counting a finite number of injections, there is an
uncertainty in the efficiency attributable to binomial counting
fluctuations.  Second, there is an uncertainty in the amplitude to
which a measurement of the efficiency should be assigned, on account
of it having been found by counting injections spanning a range of
amplitudes.  Finally, uncertainties in the calibration translate into
an additional uncertainty in the amplitude, on account of the
injections from which the efficiency was measured having been done at
amplitudes different from what was intended.  The calibration
uncertainty we use is 11\%. This number results from a the systematics
in the calibration models (5\%) and our use of time domain calibrated
data (10\%), which we combine in quadrature. The counting, amplitude
range, and calibration uncertainties described above above are
combined in quadrature to produce the shaded regions shown in
Fig.~\ref{fig:effH1}.

\section{Parameter Space of cosmic string models: Constraints and Sensitivity}
\label{cosmo}

For simplicity in this section we will adopt units where the speed of
light $c=1$. The parameter space of cosmic string models we need to
consider depends on whether loops in the cosmic string network are
short-lived (lifetime much smaller than a Hubble time) or long-lived
(lifetime much larger than a Hubble time). This, in turn, depends on
loop sizes at formation.  If their size is given by the gravitational
back reaction scale~\cite{Siemens:2002dj,Polchinski:2007rg}, then to a
good approximation all loops have the same size at formation and they
are short lived. In this case their size at formation at cosmic time
$t$ can be approximated by $l=\varepsilon \Gamma G \mu t$, where
$\varepsilon < 1$~\cite{DV2} is an unknown parameter that depends on
the spectrum of perturbations on cosmic strings, and $\Gamma$ is a
constant related to the lifetime of loops, and is measured in
simulations to be $\Gamma \sim 50$. Recent cosmic string network
simulations, however, suggest loops form at much larger sizes given by
the network dynamics~\cite{Vanchurin:2005pa,Olum:2006ix}. If this is
the case it has been shown~\cite{SMC} that the regions of parameter
space accessible to Initial LIGO are already ruled out by pulsar
timing experiments.  So here we will consider only the first
possibility, that loop sizes are determined by gravitational back
reaction and take the size of loops at formation to be $l=\varepsilon
\Gamma G \mu t$.

Unlike field theoretic cosmic strings, cosmic superstrings do not
always reconnect when they meet. Rather, they reconnect with
probability $p$ which has a value in the range
$10^{-3}$-$1$~\cite{Jackson:2004zg}. The reasons for this are (1) that
fundamental strings interact probabilistically, and (2) that in more than
3 dimensions, even though two strings may appear to meet in 3 dimensions
they miss each other in the extra dimensions. The effect of the
decreased reconnection probability is to increase the
density of strings by a factor inversely proportional to the
reconnection probability~\cite{DV2}.

For a point in cosmic string parameter space $(G\mu,\varepsilon,p)$ 
we can use the efficiency curves $\epsilon(B)$ to compute the rate of bursts we
expect to observe in our instruments, which we will refer to as the effective
rate $\gamma$. It is given by the integral~\cite{SCMMCR}
\begin{equation}
\gamma(G\mu,\varepsilon,p) = \int_0^{\infty} \epsilon (B) \frac{dR(B;G\mu,\varepsilon,p)}{dB}dB,
\label{eq:effrate}
\end{equation}
where $B\sim G\mu l^{2/3}/r$ is the optimally oriented amplitude (i.e.
the amplitude of events excluding antenna pattern effects), $\epsilon
(B)$ is the efficiency of detecting events at an amplitude $B$ and
$dR(B;G\mu,\varepsilon,p)$ is the cosmological rate of events with
amplitude in the interval $B$ and $B+dB$.  We have take the size of
the feature that produces the cusp to be the size of the loop $l$.

Since we are considering loops that are small when they are formed
they are also short lived, and at a given redshift they are all of
essentially the same size. As a result the amplitude of burst events
from a given redshift is the same. In this case, rather than
Eq.~(\ref{eq:effrate}) it is easier to evaluate,
\begin{equation}
\gamma(G\mu,\varepsilon,p) = \int^{\infty}_0 \epsilon (z) \frac{dR(z;G\mu,\varepsilon,p)}{dz}dz.
\label{eq:effrate2}
\end{equation}
Here $dR(z)$ the rate of bursts originating at redshifts in the
interval between $z$ and $z+dz$. The rate is given by Eq.~(59) of
\cite{SCMMCR}
\begin{eqnarray}
\frac {dR}{dz} &=& H_0 \frac{N_c(g_2 f_* H^{-1}_0)^{-2/3}}{2\alpha^{5/3} p
\Gamma G \mu} \varphi^{-14/3}_t(z) \varphi_V(z) (1+z)^{-5/3}
\nonumber
\\
&\times&  \Theta(1-\theta_m(z,f_*,\alpha H^{-1}_0 \varphi_t (z)),
\label{eq:rate12}
\end{eqnarray}
where $H_0$ is the present value of the Hubble parameter, $N_c$ is the
average number of cusps per loop oscillation; $g_2$ is an ignorance
constant that absorbs the unknown fraction of the loop length that
contributes to the cusp and other factors of $O(1)$; $f_*$ is the
lowest high frequency cutoff of the bursts we are interested in
detecting; $\alpha = \varepsilon \Gamma G \mu$ is the loop formation size
in units of the cosmic time; and $\theta_m=[g_2(1+z)f_*l]^{-1/3}$ is the
maximum angle a cusp and the line of sight can subtend and still
produce a burst with high frequency cutoff $f_*$.  The
$\Theta$-function removes events that don't have the form of
Eq.~(\ref{eq:hf1}). Two dimensionless cosmological functions enter the
expression for the rate of events: $\varphi_t(z)$ which relates the
cosmic time $t$ and the redshift via $t=H^{-1}_0 \varphi_t(z)$, and
$\varphi_V(z)$ which determines the proper volume element at a
redshift $z$ through $dV(z)=H^{-3}_0\varphi_V(z)dz$ (see Appendix A of
\cite{SCMMCR}). For details on the derivation of this expression 
see~\cite{DV0,DV1,DV2,SCMMCR}.

The efficiency $\epsilon (z)$ is the fraction of events we detect from
a redshift $z$. We compute this quantity starting from our measured
efficiency as a function of the amplitude $B$, using Eq.~(60) of
\cite{SCMMCR}
\begin{equation}
\frac{\varphi^{2/3}_t (z)}{(1+z)^{1/3} \varphi_r (z)} = \frac{B H^{-1/3}_0
}{g_1 G\mu \alpha^{2/3}},
\label{eq:zofA1}
\end{equation}
where $g_1$ is an ignorance constant that absorbs the unknown fraction
of the loop length that contributes to the cusp and factors of $O(1)$
(different from those of $g_2$, see~\cite{SCMMCR} for details), and
$\varphi_r(z)$ is a dimensionless cosmological function that relates
the proper distance $r$ to the redshift via $r=H^{-1}_0 \varphi_r(z)$.
Solving Eq.~(\ref{eq:zofA1}) for $z$ gives the redshift from which a
burst of amplitude $B$ originates. Thus for each amplitude in our
efficiency curve we can determine the corresponding redshift and
construct $\epsilon(z)$.

\begin{figure*}[hbtp]
\includegraphics[width=3.5in]{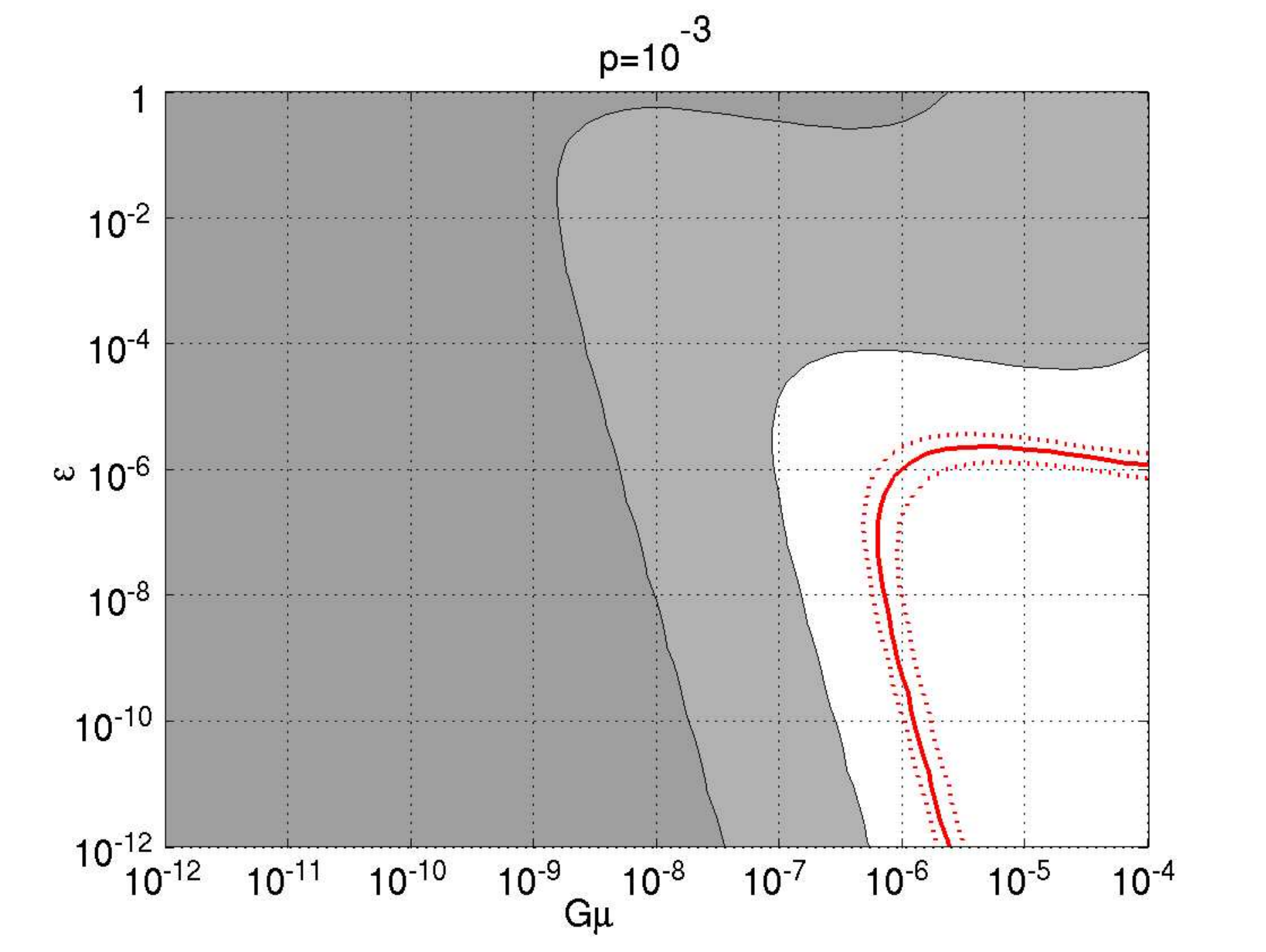}
\includegraphics[width=3.5in]{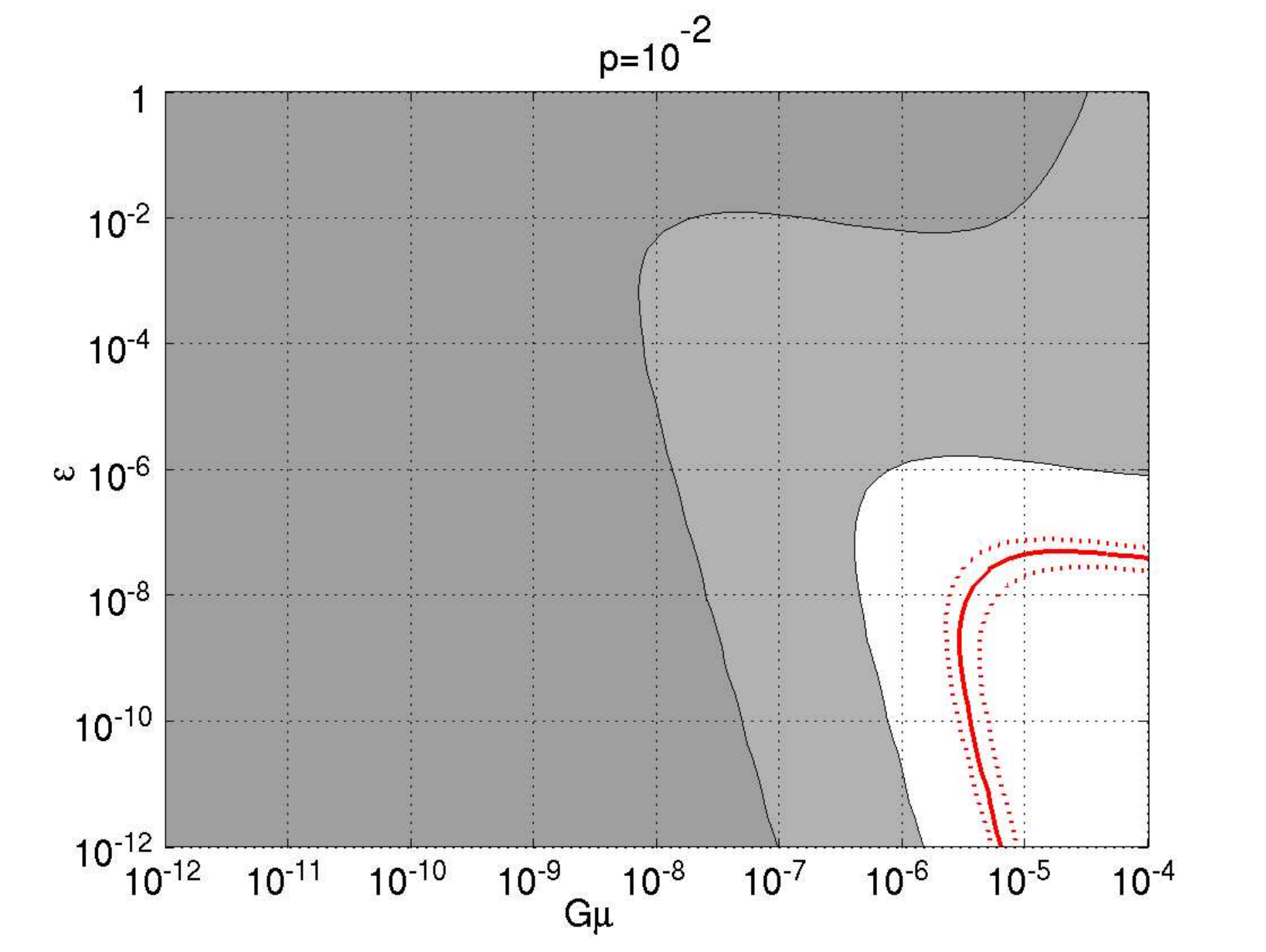}
\includegraphics[width=3.5in]{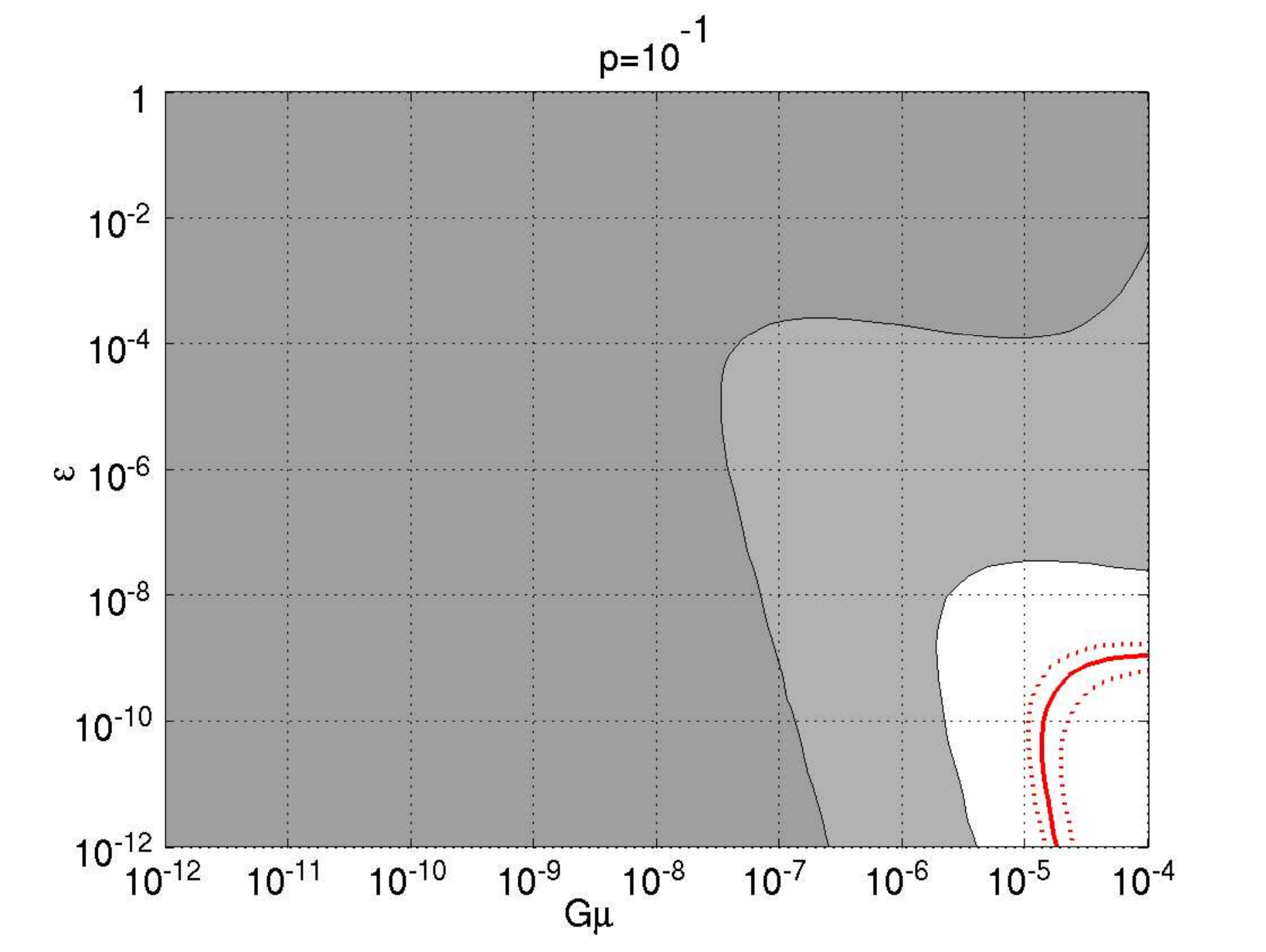}
\includegraphics[width=3.5in]{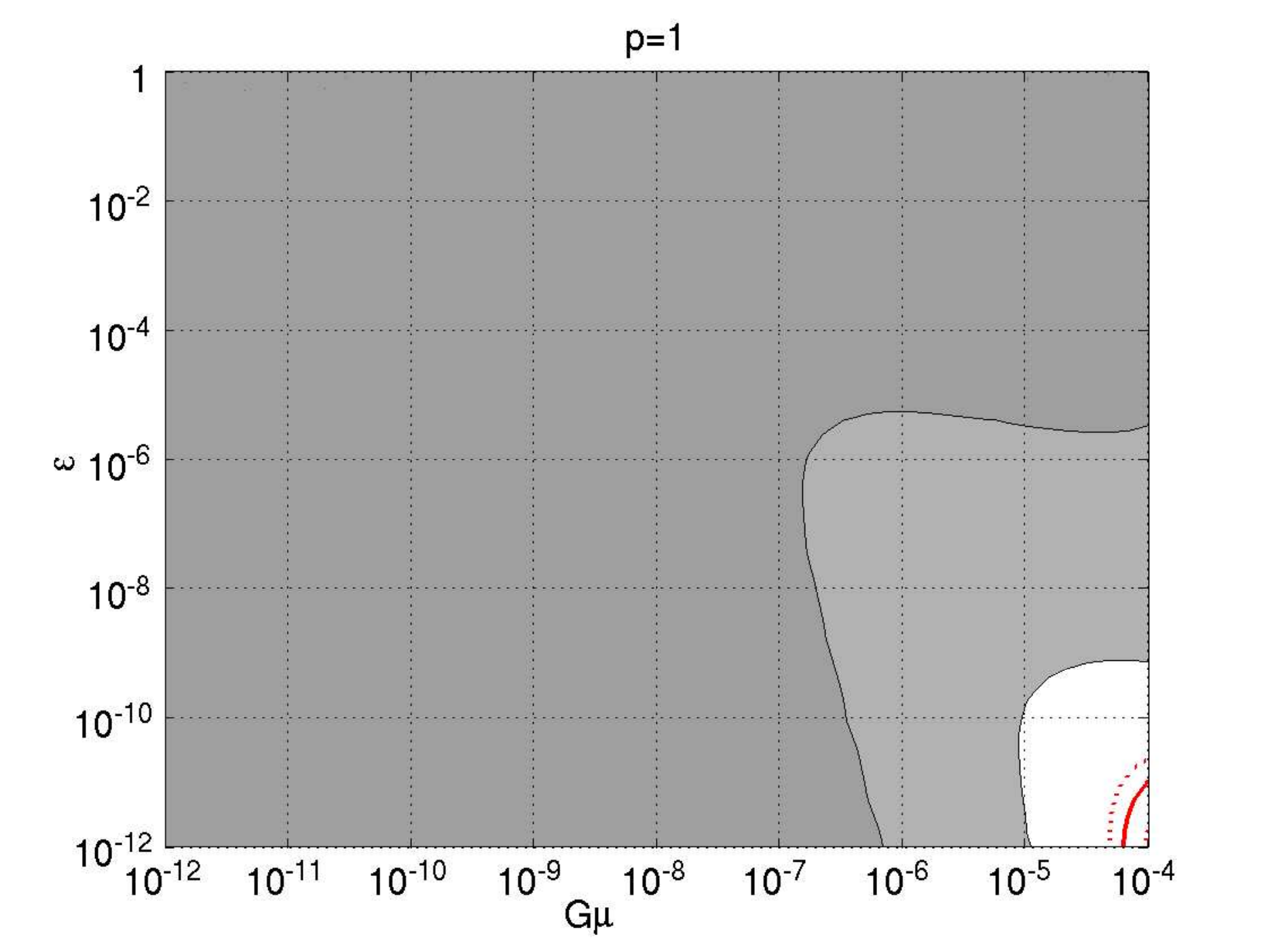}
\caption{Plot of the sensitivity and upper limit results of our
  analysis. Each of the four panels corresponds to a value of the
  reconnection probability $p$ with $y$-axes $\varepsilon$ and
  $x$-axes $G\mu$. Areas to the right of the red curves show the
  regions excluded at the 90\% level by our analysis. The dotted
  curves indicate the uncertainty in the areas of parameter space
  excluded that arise from uncertainties in the efficiency. The light
  and dark gray shaded areas are regions of parameter space unlikely
  to result in a cosmic string cusp event detected in S4: a cosmic
  string network with model parameters in these regions would result
  in less than one event (on average) surviving our pipeline.  The
  lower boundary of these shaded areas was computed using the efficiency for
  all recovered injections, the dashed curve shown in Fig.~\ref{fig:effH1}. The
  dark gray regions show regions of parameter space unlikely to result
  in a cosmic string cusp being detected in a year long search of
  Initial LIGO data.  The lower boundary of the dark gray areas was computed with the
  Initial LIGO sensitivity estimate $B^{{\rm LIGO}}_{50\%} \approx
  10^{-20}$~s$^{-1/3}$~\cite{SCMMCR}.}
\label{results1}
\end{figure*}

We have computed the cosmological functions $\varphi_r(z)$,
$\varphi_t(z)$, and $\varphi_V(z)$, as described in Appendix A
of~\cite{SCMMCR}, using the latest consensus cosmological parameters
measured by WMAP~\cite{wmap}. Specifically, we used a present day
Hubble parameter of $H_0 = 70.1\,$km$\,$~s$^{-1}$Mpc$^{-1}$ $= 2.27 \times
10^{-18}\rm{s}^{-1}$, and densities relative to the critical density
$\Omega_m = 0.279$ for matter, $\Omega_r = 8.5 \times 10^{-5}$ for
radiation, and $\Omega_\Lambda=1-\Omega_m-\Omega_r=0.721$ for the
cosmological constant.

We scan the parameter space of the small loop models by varying the
reconnection probability $p$, the dimensionless string tension $G\mu$,
and the size of loops parametrized by $\varepsilon$.  To construct a
90\% upper limit we use the loudest event
statistic~\cite{loudestevent}. For each point in cosmic string
parameter space we compute the effective rate $\gamma$ using the solid
black curve in Fig.~\ref{fig:effH1} (recovered injections with
amplitudes above our loudest event). We then compare this rate with
$\gamma_{90\%}=2.303/T$, the rate for which 90\% of the time we would
have seen at least one event in a Poisson process if we observed for a
time $T$.  If for a point in parameter space the effective rate
$\gamma(G\mu, \varepsilon,p)$ exceeds $\gamma_{90\%}$ then we say
those parameters are ruled out at the 90\% level, in the sense that
cosmic string models with those parameters would have produced an
event with amplitude larger than our loudest event 90\% of the time.
We can also estimate our sensitivity by computing $\gamma$ using the
dashed curve in Fig.~\ref{fig:effH1} (recovered injections with any
amplitude) and then compare it to $1/T$. The latter tells us what
models would have resulted in at least one event (on average)
surviving our pipeline, a useful measure of our sensitivity. For our
search in triple coincident S4 data $T=14.9$ days.

Figure~\ref{results1} shows the results of our analysis. We have set
$N_c=g_1=g_2=1$ for convenience. Each of the four panels corresponds
to a value of the reconnection probability $p$ with the loop size
parameter $\varepsilon$ on the $y$-axis, and the dimensionless string
tension $G\mu$ on the $x$-axis. Regions to the right of the red curves
are excluded at the 90\% level by our analysis. The dotted curves show
the uncertainty in the areas of parameter space excluded that result
from uncertainties in the efficiency.  The light and dark gray shaded
areas in each of the four panels are regions of parameter space very
unlikely to result in a cosmic string cusp event detected in S4: a
cosmic string network with model parameters in these regions would
result in less than one event (on average) surviving our pipeline.
The lower boundary of these areas was computed using the efficiency
for all injections, the dashed curve shown in Fig.~\ref{fig:effH1},
and comparing the effective rate to $1/T$.  The dark gray regions show
regions of cosmic string model parameter space unlikely to result in a
cosmic string cusp being detected in a year long search of Initial
LIGO data: on average such models would result in fewer than one event
being detected.  The lower boundary curve of the dark gray regions was
computed with the Initial LIGO sensitivity estimate $B^{{\rm
 LIGO}}_{50\%} \approx 10^{-20}$~s$^{-1/3}$~\cite{SCMMCR} and
assuming a year of observation time.

Due to the sensitivity and duration of the S4 run the 90\% limits we
have placed on the parameter space of cosmic string models are not as
constraining as the indirect bounds due to Big Bang
Nucleosynthesis~\cite{SMC}. Another current gravitational wave bound
comes from pulsar timing observations. Due to their sensitivity at
very low frequencies pulsar timing bounds for loop sizes given by
gravitational back reaction constrain an independent portion of the
cosmic string parameter space~\cite{SMC}.  Future analysis of data from the
fifth science run, however, a factor of two more sensitive and with a
year of triple coincident data, will be sufficiently sensitive to
surpass these limits in large areas of cosmic string parameter
space.

\section{Summary}
\label{concl}

We have performed a search for bursts from cosmic string cusps in 14.9
days of triple coincident data from LIGO's fourth science run. The
gravitational waveforms of cosmic string cusps are known and
matched-filters provide the optimal means of extracting such signals
from noisy data. We constructed a template bank and generated a set of
triggers for each of the three LIGO interferometers. To reduce the
rate of accidentals we demanded the resulting triggers satisfy two
consistency criteria: (i) time coincidence between events in the three
instruments, and (ii) amplitude consistency between events in the two
Hanford interferometers H1 and H2.  The latter check is possible
because the Hanford instruments are co-aligned.  The effect of the
consistency criteria is to reduce the trigger rate in each of the
instruments from about 1~Hz to about 10~$\mu$Hz.  To estimate our
background, the rate of accidentals, we performed 100 time-slides on
the data, with a time step much larger than the duration of our
signals.  Comparing our background estimate with our foreground, we
conclude no gravitational waves have been found in this search.

To estimate the sensitivity of the search and place constraints on the
parameter space of cosmic strings we performed several thousand
simulated signal injections into our data streams and attempted to
recover them. The simulated signal parameters used are consistent with
our expectations for the cosmic string population. The injections were
used to compute the efficiency, the fraction of events that we detect
in a range of amplitudes. The efficiency curves can be convolved with
the cosmological rate of events to compute the so-called effective
rate: the rate of detectable events in our search.  The effective rate
folds together the properties of the population, such as the
distribution of sources in the sky, with the sensitivity of the
detectors and the analysis pipeline. We found our estimate for the
sensitivity of this search to be consistent with our expectations,
given that we used S4 data and included H2 in the analysis. Using the
loudest event in our foreground, we placed a frequentist 90\% upper
limit on the effective rate, which we in turn used to constrain the
parameter space of cosmic strings models.  Unfortunately, the
sensitivity of this search does not allow us to place constraints as
tight as the indirect bounds from Big Bang Nucleosynthesis.  However,
analyses using data from future LIGO runs is expected to surpass these
limits for large areas of cosmic string parameter space.

\acknowledgements
The authors would like to thank Irit Maor and Alexander Vilenkin for
useful discussions. The authors gratefully acknowledge the support of
the United States National Science Foundation for the construction and
operation of the LIGO Laboratory and the Particle Physics and
Astronomy Research Council of the United Kingdom, the
Max-Planck-Society and the State of Niedersachsen/Germany for support
of the construction and operation of the GEO600 detector. The authors
also gratefully acknowledge the support of the research by these
agencies and by the Australian Research Council, the Natural Sciences
and Engineering Research Council of Canada, the Council of Scientific
and Industrial Research of India, the Department of Science and
Technology of India, the Spanish Ministerio de Educacion y Ciencia,
The National Aeronautics and Space Administration, the John Simon
Guggenheim Foundation, the Alexander von Humboldt Foundation, the
Leverhulme Trust, the David and Lucile Packard Foundation, the
Research Corporation, and the Alfred P. Sloan Foundation.  LIGO DCC
number: LIGO-P0900026.


\begin{thebibliography}{}

\bibitem{kibble76}
T.~W.~B. Kibble,  \emph{J. Phys.} \textbf{A9} 1387, 1976.

\bibitem{alexbook}
A.~Vilenkin and E.~Shellard, Cosmic strings and other Topological Defects
  (Cambridge University Press, 2000).

\bibitem{Jeannerot:2003qv}
R.~Jeannerot, J.~Rocher, and M.~Sakellariadou, \emph{Phys. Rev.} \textbf{D68} 103514, 2003.

\bibitem{Jones:2002cv}
N.~T. Jones, H.~Stoica, and S.~H.~H. Tye, \emph{JHEP} \textbf{07} 051, 2002.

\bibitem{Sarangi:2002yt}
S.~Sarangi and S.~H.~H. Tye,  \emph{Phys. Lett.} \textbf{B536} 185, 2002.

\bibitem{Dvali:2003zj}
G.~Dvali and A.~Vilenkin, \emph{JCAP} \textbf{0403} 010, 2004.

\bibitem{Jones:2003da}
N.~T. Jones, H.~Stoica, and S.~H.~H. Tye, \emph{Phys. Lett.}
  \textbf{B563} 6, 2003.

\bibitem{Copeland:2003bj}
E.~J. Copeland, R.~C. Myers, and J.~Polchinski,  \emph{JHEP} \textbf{06} 013, 2004.

\bibitem{Jackson:2004zg}
M.~G. Jackson, N.~T. Jones, and J.~Polchinski,\emph{JHEP} \textbf{10} 013, 2005.

\bibitem{Tye:2005fn}
S.~H.~H. Tye, I.~Wasserman, and M.~Wyman, \emph{Phys. Rev.} \textbf{D71} 103508, 2005.

\bibitem{Copeland:2006if}
E.~J. Copeland, T.~W.~B. Kibble, and D.~A. Steer, \emph{Phys. Rev.} \textbf{D75} 065024, 2007.

\bibitem{Leblond:2007tf}
L.~Leblond and M.~Wyman,  \emph{Phys. Rev.}
  \textbf{D75} 123522, 2007.

\bibitem{Rajantie:2007hp}
A.~Rajantie, M.~Sakellariadou, and H.~Stoica, 2007.

\bibitem{berez}
  V.~Berezinsky, B.~Hnatyk and A.~Vilenkin,
  Phys.\ Rev.\  D {\bf 64}, 043004 (2001)
  [arXiv:astro-ph/0102366].

\bibitem{bands}
  P.~Bhattacharjee and G.~Sigl,
  Phys.\ Rept.\  {\bf 327}, 109 (2000)
  [arXiv:astro-ph/9811011].


\bibitem{Chernoff:2007pd}
  D.~F.~Chernoff and S.~H.~H.~Tye,
  arXiv:0709.1139 [astro-ph].

\bibitem{Kuijken:2007ma}
  K.~Kuijken, X.~Siemens and T.~Vachaspati,
  arXiv:0707.2971 [astro-ph].

\bibitem{Gasparini:2007jj}
  M.~A.~Gasparini, P.~Marshall, T.~Treu, E.~Morganson and F.~Dubath,
  arXiv:0710.5544 [astro-ph].

\bibitem{Christiansen:2008vi}
  J.~L.~Christiansen, E.~Albin, K.~A.~James, J.~Goldman, D.~Maruyama and G.~F.~Smoot,
  Phys.\ Rev.\  D {\bf 77}, 123509 (2008)
  [arXiv:0803.0027 [astro-ph]].

\bibitem{Dyda:2007su}
  S.~Dyda and R.~H.~Brandenberger,
  arXiv:0710.1903 [astro-ph].


\bibitem{Vachaspati:2008su}
  T.~Vachaspati,
  arXiv:0802.0711 [astro-ph].

\bibitem{Khatri:2008zw}
  R.~Khatri and B.~D.~Wandelt,
  Phys.\ Rev.\ Lett.\  {\bf 100}, 091302 (2008)
  [arXiv:0801.4406 [astro-ph]].

\bibitem{Battefeld:2007qn}
  D.~Battefeld, T.~Battefeld, D.~H.~Wesley and M.~Wyman,
  JCAP {\bf 0802}, 001 (2008)
  [arXiv:0708.2901 [astro-ph]].

\bibitem{Fraisse:2007nu}
  A.~A.~Fraisse, C.~Ringeval, D.~N.~Spergel and F.~R.~Bouchet,
  Phys.\ Rev.\  D {\bf 78}, 043535 (2008)
  [arXiv:0708.1162 [astro-ph]].

\bibitem{Pogosian:2008am}
  L.~Pogosian, S.~H.~Tye, I.~Wasserman and M.~Wyman,
  arXiv:0804.0810 [astro-ph].

\bibitem{Baumann:2008aj}
  D.~Baumann {\it et al.},
  arXiv:0811.3911 [astro-ph].

\bibitem{firstcusps}  V. Berezinsky et al., astro-ph/0001213.

\bibitem{DV0} T. Damour and A. Vilenkin, Phys. Rev. Lett. 85 (2000) 3761.

\bibitem{DV1} T. Damour and A. Vilenkin, Phys. Rev. D 64 (2001) 064008.

\bibitem{DV2} T. Damour and A. Vilenkin, Phys. Rev. D 71 (2005) 063510.

\bibitem{SCMMCR} X. Siemens et al., Phys. Rev {\bf D}73 (2006) 105001.

\bibitem{ligoref} A.~Abramovici et al., Science {\bf 256}, 325 (1992);
B.~Barish and R.~Weiss, Phys. Today {\bf 52}, 44 (1999).

\bibitem{virgoref} B.~Caron et al., Nucl. Phys. B-Proc. Suppl. 
{\bf 54}, 167 (1997).

\bibitem{kenandi} X. Siemens and K.D. Olum, Phys.Rev. {\bf D}68 (2003) 085017.

\bibitem{damour} D. Chialva and T. Damour, hep-th/0606226.


\bibitem{sigg} D.~Sigg (for the LIGO Scientific Collaboration),
  Class. Quantum Grav. {\bf 23} (2006) S51.

\bibitem{loudestevent} P.R. Brady et al. Class.Quant.Grav.{\bf 21} (2004) S1775.

\bibitem{SMC} X.~Siemens, V.~Mandic and J.~Creighton,
Phys.\ Rev.\ Lett.\  {\bf 98}, 111101 (2007) [arXiv:astro-ph/0610920].

\bibitem{htpaper} X. Siemens et al., Class.Quant.Grav. {\bf 21} (2004) S1723.

\bibitem{cutlerflanagan}  C. Cutler and E.E. Flanagan, Phys.Rev.{\bf
    D} 49 (1994) 2658.

\bibitem{Allen:2005fk}
  B.~Allen, W.~G.~Anderson, P.~R.~Brady, D.~A.~Brown and J.~D.~E.~Creighton, arXiv:gr-qc/0509116.

\bibitem{Siemens:2002dj}
  X.~Siemens, K.~D.~Olum and A.~Vilenkin,
  Phys.\ Rev.\  D {\bf 66}, 043501 (2002)
  [arXiv:gr-qc/0203006].

\bibitem{Polchinski:2007rg}
  J.~Polchinski and J.~V.~Rocha,
  Phys.\ Rev.\  D {\bf 75}, 123503 (2007)
  [arXiv:gr-qc/0702055].

\bibitem{Vanchurin:2005pa}
  V.~Vanchurin, K.~D.~Olum and A.~Vilenkin,
  Phys.\ Rev.\  D {\bf 74}, 063527 (2006)
  [arXiv:gr-qc/0511159].

\bibitem{Olum:2006ix}
  K.~D.~Olum and V.~Vanchurin,                                
  Phys.\ Rev.\  D {\bf 75}, 063521 (2007)
  [arXiv:astro-ph/0610419].

\bibitem{wmap} http://lambda.gsfc.nasa.gov/product/map/dr3/parameters\_summary.cfm
  
\end{thebibliography}
\end{document}